\documentclass[preprint,onecolumn,showpacs,nobibnotes,epsf,amsmath,amssymb]{revtex4}

\usepackage{graphicx}
\usepackage{dcolumn}
\usepackage{bm}
\usepackage{SIunits}

\begin{document}

\title{Abnormal T-linear susceptibility and Phase diagram of BaFe$_{2-x}$Co$_x$As$_2$ single crystals}
\author{X. F. Wang, T. Wu, G. Wu, R. H. Liu, H. Chen, Y. L. Xie and X. H. Chen$^\ast$} \affiliation{Hefei National
Laboratory for Physical Science at Microscale and Department of
Physics, University of Science and Technology of China, Hefei, Anhui
230026, People's Republic of China}

\begin{abstract}
We study systematically transport, susceptibility and heat capacity
for BaFe$_{2-x}$Co$_x$As$_2$ single crystals. In the underdoped
region, spin density wave (SDW) transition is observed in both
resistivity and susceptibility. The magnetic susceptibility shows
unusual T-linear dependence above SDW transition up to 700 K. With
Co doping, SDW ordering is gradually suppressed and
superconductivity emerges with a dome-like shape. Electrical
transport, specific heat and magnetic susceptibility indicate that
SDW and superconductivity coexist in the sample
BaFe$_{2-x}$Co$_x$As$_2$ around x = 0.17, being similar with
(Ba,K)Fe$_2$As$_2$. When x$>$0.34, the superconductivity completely
disappears. A crossover from non-Fermi-liquid state to Fermi-liquid
state is observed with increasing Co doping. A detailed electronic
phase diagram about evolution from SDW to superconducting state is
given.
\end{abstract}


\maketitle

\section{Introduction}

Recent discovery of iron-based arsenide high-Tc
superconductor\cite{yoichi, chenxh, chen, ren, rotter} attracts much
interesting. Up to now, the pnictide superconductors with two main
kinds of crystallographic structure have been widely studied:
ZrCuSiAs-type (1111) and ThCr$_2$Si$_2$-type (122). Similar to
cuprates, such pnictide superconductors are also believed to have a
quasi-2D conducting layer---Fe$_2$As$_2$ layer, which is separated
by LnO (Ln = La, Sm etc.) or R (R = Ba, Sr etc.) charge reservior.
Electron and hole can be introduced to FeAs layer to realize
superconductivity by replacing elements in charge
reservior\cite{yoichi, dong, liu, Luetkens, zhao, Drew, chenhong}.
However, in contrast to high-Tc cuprates, doping by replacing Fe
atom with Co or Ni atom in FeAs layer can also produce
superconductivity in both 1111 and 122 structures\cite{sefat, cwang,
alj, sefat2, sefat3, Kumar, Ahilan}. For example, in Co-doped
BaFe$_2$As$_2$, the superconducting temperature reach up to 22 K,
lower than 38 K in K-doped BaFe$_2$As$_2$\cite{sefat, rotter}. It
seems that superconductivity in FeAs layer is very robust, and the
integrality of FeAs layer is not important for superconductivity,
being in sharp contrast to the case of cuprates in which any doping
in Cu site destroyed the superconductivity. It makes the origin of
superconductivity in iron-arsenide very complicated. However,
detailed study on Co-doping system is very limit\cite{sefat, cwang,
alj, sefat2, sefat3, Kumar, Ahilan}. Systematic study on Co-doped
system and phase diagram are urgent needs to elucidate the role of
Co-doping and to understand the superconductivity. In this paper,
systematic study on transport, magnetism and heat capacity of
BaFe$_{2-x}$Co$_x$As$_2$ single crystals are reported. In the
underdoped region, spin density wave (SDW) transition is observed in
both transport and magnetism. Interestingly, the magnetic
susceptibility shows unusual T-linear dependence above SDW
transition up to 700 K. It indicates strong antiferromagnetic
coupling above SDW ordering in this system. With Co doping, SDW
transition is suppressed and superconducting state gradually emerges
with a dome-like shape. A crossover from non-Fermi-liquid to
Fermi-liquid behavior is evidenced by transport. The coexistence of
SDW and superconductivty is observed around x=0.17. When x$>$0.34,
the superconducting state completely disappears and T-linear
susceptibility gradually changes to Curie-weiss law. A detailed
electronic phase diagram about evolution from SDW to superconducting
state is presented.

\section{Experiment}

Single crystals of $BaFe_{2-x}Co_{x}As_{2}$ were grown by self-flux
method. In order to avoid contamination from incorporation of other
elements into the crystals, FeAs was chosen as the
flux\cite{xfwang}. FeAs and CoAs powder was mixed together, then
roughly grounded. The Ba pieces were added into the mixture. All
procedures above are achieved in glove box in which high pure argon
atmosphere is filled. The total proportion of Ba:(2-xFeAs+xCoAs) is
1:4. The details have been reported elsewhere\cite{xfwang}. Single
crystals were characterized by x-ray diffractions (XRD) using Cu
$K_{\alpha}$ radiations. As shown in Fig. 1, only (00l) peaks is
observed, indicating that the single crystals are perfect
c-orientation. The parameter of the c-axes linearly decreases with
increasing Co-doping, being consistent with the results previously
reported\cite{liyk}. The actual chemical composition of the single
crystals is determined by inductively coupled plasma (ICP) atomic
emission spectroscopy (AES) (ICP-AES) technique and X-ray Energy
Dispersive Spectrum (EDS). The electrical transport was measured
using the ac four-probe method with an alternative current (ac)
resistance bridge system (Linear Research, Inc.; LR-700P). Hall
effect is measured by four-terminal ac technique. The magnetic
susceptibility was measured by SQUID (Quantum Design). The in-plane
magnetoresistivity was measured by Physical Properties Measurement
System (PPMS, Quantum Design).

\section{Results and Discussions}

Fig.2 shows the temperature dependence of in-plane and out-of-plane
resistivity for BaFe$_{2-x}$Co$_x$As$_2$ (0.08$\sim$x$\sim$0.60).
For x $<$ 0.18, an upturn behavior in ab-plane resistivity is
clearly observed. With Co doping, the upturn temperature decreases
and superconductivity emerges at x=0.17. In BaFe$_2$As$_2$, this
abnormal behavior in resistivity is ascribed to SDW
transition/structural transition\cite{rotter}. Here, the upturn
behavior is also believed to originate from SDW transition or
structural transition. For out-of-plane resistivity, a similar
behavior is also observed. It is very surprising that the SDW
transition seems to coexist with superconductivity for the samples
with x = 0.17 and x = 0.18. Similar results is also observed in
K-doped BaFe$_2$As$_2$ system\cite{chenh}. This issue will be
further discussed later. With further doping, the upturn behavior is
completely suppressed and superconductivity can reach to the maximum
25 K at x = 0.2 sample, being consistent with the previous
report\cite{sefat}. The superconducting  transition temperature
decreases with further doping, and superconductivity disappears for
the samples with x$>$0.34. As shown in Fig. 2(c), the anisotropic
ratio monotonously decreases with Co doping. A weak temperature
dependence for the anisotropy ($\rho_c/\rho_c$) is observed for all
doping. It indicates that in-plane and out-of-plane transport shares
the same scattering mechanism, and interplane correlation is
enhanced with Co doping, being consistent with the fact that the
parameter of c-axis decreases with doping Co. We also fit
resistivity in the whole temperature range for the samples without
SDW ordering by the formula: R=A+BT$^{n}$. The values of parameter n
are listed in Table I. The value of n increases from 1.25 for
underdoped sample with x=0.18 to 2.01 for overdoped sample with
x=0.38. It is well known that the T-linear behavior is believed to
arise from non-Fermi-liquid system as observed in curaptes, while
T$^2$ behavior originates from Fermi-liquid system. These results
indicate that this system seems to show a crossover from
non-Fermi-liquid state to Fermi-liquid state.

Fig. 3 shows the field-cooling (FC) and zero-field-cooling (ZFC)
magnetic susceptibility for the superconducting samples in the
temperature range from 2 K to 30 K at a fixed magnetic field 5 Oe.
All the samples show a sharp superconductivity transition. We have
estimated the superconducting volume by ZFC susceptibility. For the
samples with x = 0.17,0.25 and 0.18, the superconductivity volume
fraction reaches to 100\% at 2 K. For x = 0.20, the volume is reach
to 80\% at 2 K. It indicates a bulk superconductivity in these
samples.

Fig. 4 shows the temperature dependent in-plane susceptibility for
all samples in the whole temperature range under H = 6.5 T. As shown
in Fig. 3(a), a T-linear behavior is observed for the samples with
SDW ordering above SDW temperature. The value of room-temperature
susceptibility decreases with Co doping. But the slope does not
change with Co doping for the samples with x $<$ 0.16. For x = 0.17,
the slope begins to decrease, but T-linear behavior keeps. For x =
0.2, T-linear behavior is broken and an upturn behavior is observed
in low temperature region. In order to investigate the T-linear
behavior, we have expanded the temperature range up to 700 K. As
shown in Fig.3(b), a well-defined T-linear is observed for x = 0 and
0.17 samples up to 700 K. Recently, Zhang et al. have given an
explanation to understand the T-linear behavior in
susceptibility\cite{zhang}. These data definitely indicates a strong
antiferromagnetic coupling above the SDW transition. Strong magnetic
fluctuation is believed to be the key point. Here, a T-linear
behavior up to 700 K indicates that strong magnetic fluctuation
exists in this system, and it remains in supercconducting samples.
As shown in Fig.3(c), for overdoped sample, the deviation from
T-linear behavior becomes more obvious and a Curie-weiss like
behavior is observed for x = 0.6 samples. It indicates that
superconductivity is somewhat related to strong antiferromagnetic
correlation. Interestingly, both T-linear behavior and
superconductivity disappear at the meantime.

Fig. 5 is temperature-dependent Hall coefficient for the crystals
$BaFe_{2-x}Co_{x}As_{2}$ (x=0.08, 0.16, 0.20, 0.60). For x = 0.08
and 0.16, the Hall coefficient shows a large slope at the
temperature corresponding to the anomaly in resistivity, being
similar to that observed in BaFe$_2$As$_2$. For x = 0.2 sample,
strong temperature-dependent behavior is observed above the
superconducting temperature. However, a weak temperature dependence
is observed for x = 0.6 sample. It is well known that
temperature-independent Hall coefficient is expected in Fermi liquid
state. Strong temperature dependence in Hall coefficient is always
ascribed to non-Fermi-liquid state, such as cuprates. The evolution
of Hall coefficient with Co doping observed here indicates that the
system evolves from non-Fermi-liquid state with Co-doping. This
result is consistent with the transport result as shown in Fig.2.

Fig. 6 shows the in-plane magnetoresistivity for the crystals
$BaFe_{2-x}Co_{x}As_{2}$ under a fixed magnetic field of 14 Tesla
rotating H within ab plane. A two-fold symmetry is observed only
below the SDW/the structure transition temperature for all the
samples with SDW ordering, and disappears above SDW ordering
temperature as shown in in Fig.6. Such two-fold symmetry disappears
for the samples without SDW ordering. In BaFe$_2$As$_2$, a similar
two-fold symmetry is also observed and it is ascribed to SDW
ordering\cite{xfwang}. These results indicate that SDW transition in
Co-doped BaFe$_2$As$_2$ has a similar magnetic symmetry to that of
BaFe$_2$As$_2$. For the samples with x=0.17 and 0.18, two-fold
symmetry also supports coexistence of superconductivity and SDW. In
order to study the coexistence of superconductivity and SDW, heat
capacity for x = 0.17 sample is studied. As shown in Fig.7, an
apparent jump is observed around SDW temperature. At low
temperatures, superconductivity also shows a peak in heat capacity
although it is very weak. In order to see clearly, we have also
measured the x = 0.17 sample under magnetic field of 14 Tesla. The
data shown in down-inset in Fig. 7 is obtained by substracting the
heat capacity of the sample measured under magnetic field of 14
Tesla. As shown in the inset, a clear jump can be seen around
superconducting temperature determined by resistivity. These results
clearly prove that the superconductivity and SDW ordering coexist in
Co-doped BaFe$_2$As$_2$ system. For x = 0.20, a clear jump is
observed in heat capacity as shown Fig. 7(b) with T$_c$ = 25.3 K. It
indicates a bulk superconductivity in this sample, being consistent
with magnetization data.

Finally, a detailed electronic phase diagram is given for the
BaFe$_{2-x}$Co$_{x}$As$_{2}$ with $0{\leqslant}x{\leqslant }0.40$.
The $T_{s}$ and $T_{c}$ was defined as the anomaly in resistivity
and the superconductivity transition by resistivity and
susceptibility, respectively. The cyan area refers to the SDW state
and the yellow area represents SC state. As been the same as the
$Ba_{1-x}K_{x}Fe_{2}As_{2}$ system, there exists a region in which
the SDW and superconductivity coexist (red area). In
BaFe$_{2-x}$Co$_{x}$As$_{2}$system ,the region is in the doping
range of $0.15{\leqslant}x{\leqslant}0.20$. It is strong evidence
that in iron-based superconducting system, the SDW and the
superconductivity coexist. In addition, these results also indicate
that a crossover form non-Fermi-liquid state to Fermi-liquid state
happens with Co doping. For underdoped samples, strong
antiferromagnetic fluctuation above SDW ordering and strong
temperature-dependent Hall coefficient are evidence for
non-Fermi-liquid state. The power law temperature dependence for the
resistivity in underdoped samples indicates that the value of n
evolves from 1 to 2 with increasing Co doping. With further
increasing Co doping, Curie-weiss like behavior and almost
temperature-independent Hall coefficient emerge. T$^2$ behavior is
fitted well in the whole temperature range for the overdoped
samples. All these results support that Fermi-liquid state emerges
with Co doping. In other hand, T-linear behavior up to 700 K in
susceptibility indicates a strong magnetic fluctuation above the SDW
ordering. Such T-linear behavior can be also observed in the
superconducting samples. It suggests that strong magnetic
fluctuation could be related to the superconductivity. This could be
very important for the superconducting mechanism.

\section{Conclusions}

Systematic study on transport, susceptibility and heat capacity for
BaFe$_{2-x}$Co$_x$As$_2$ single crystals is performed. A crossover
from non-Fermi-liquid state to Fermi-liquid state with Co-doping is
evidenced by transport and susceptibility. The magnetic
susceptibility shows unusual T-linear dependence above SDW
transition up to 700 K, indicating strong magnetic fluctuation in
this system. Electrical transport, specific heat and magnetic
susceptibility indicate that SDW and superconductivity coexist in
the sample BaFe$_{2-x}$Co$_x$As$_2$ with x = 0.17, being similar
with (Ba,K)Fe$_2$As$_2$. A detailed phase diagram with evolution
from SDW to superconducting state with Co doping is given.

 {\bf Note:} When we are preparing this manuscript, the competing
work is reported by N. Ni et al., arXiv:0811.1767 and J. H. Chu et
al., arXiv:0811.2463.

{\bf ACKNOWLEDGEMENT} This work is supported by the National Natural
Science Foundation of China and by the Ministry of Science and
Technology of China
(973 project No: 2006CB601001).\\

\vspace*{5mm} \noindent
 $^{\ast}$ Corresponding author. \emph{Electronic
address:} chenxh@ustc.edu.cn

\newpage

\begin{table}[h]
\caption{The value of parameter n fitted by power law
$\rho$=A+BT$^{n}$.}
\label{aggiungi} \centering %
\begin{ruledtabular}
\begin{tabular}{c|c|c|c|c|c|c|c}
  Co content x  & 0.18  & 0.20  & 0.25  & 0.27  & 0.32  & 0.34  & 0.38\\\hline
  n  & 1.25  & 1.27  & 1.33  & 1.55  & 1.79  & 1.90 & 2.01\\
\end{tabular}
\end{ruledtabular}
\end{table}

\begin{figure*}[t]
\includegraphics[width=12cm]{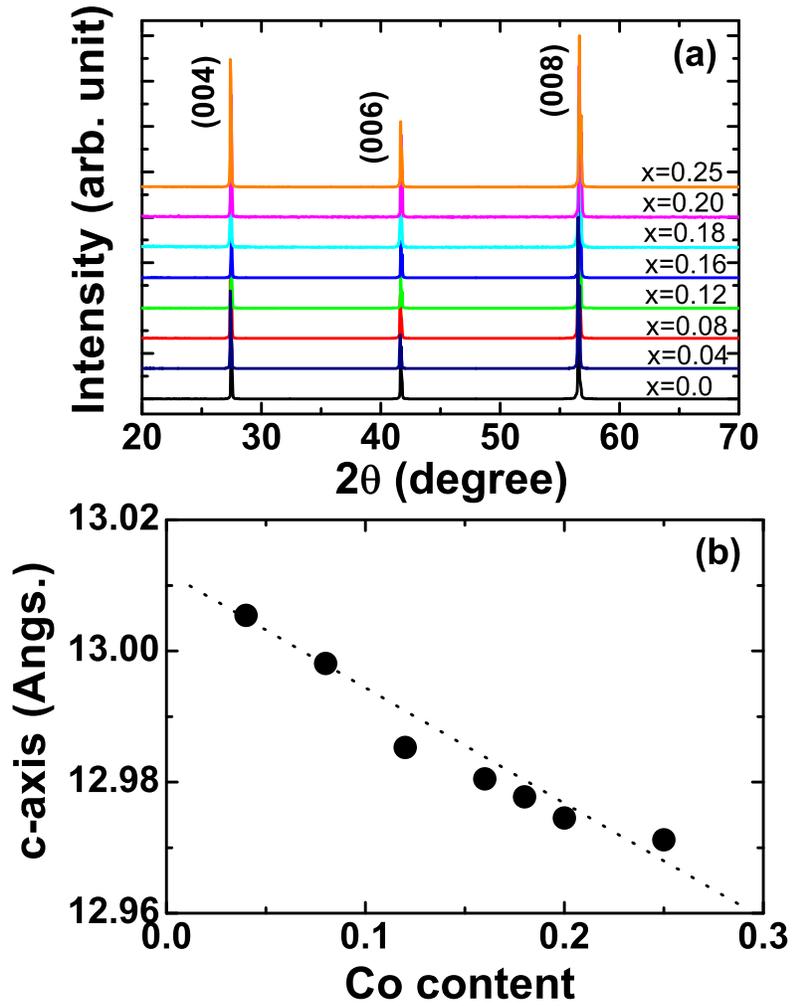}
\caption{ (a): X-ray diffraction patterns for $BaFe_
{2-x}Co_{x}As_{2}$ single crystals; (b): Doping dependence of c-axis
parameter.}
\end{figure*}

\begin{figure*}[t]
\includegraphics[width=15cm]{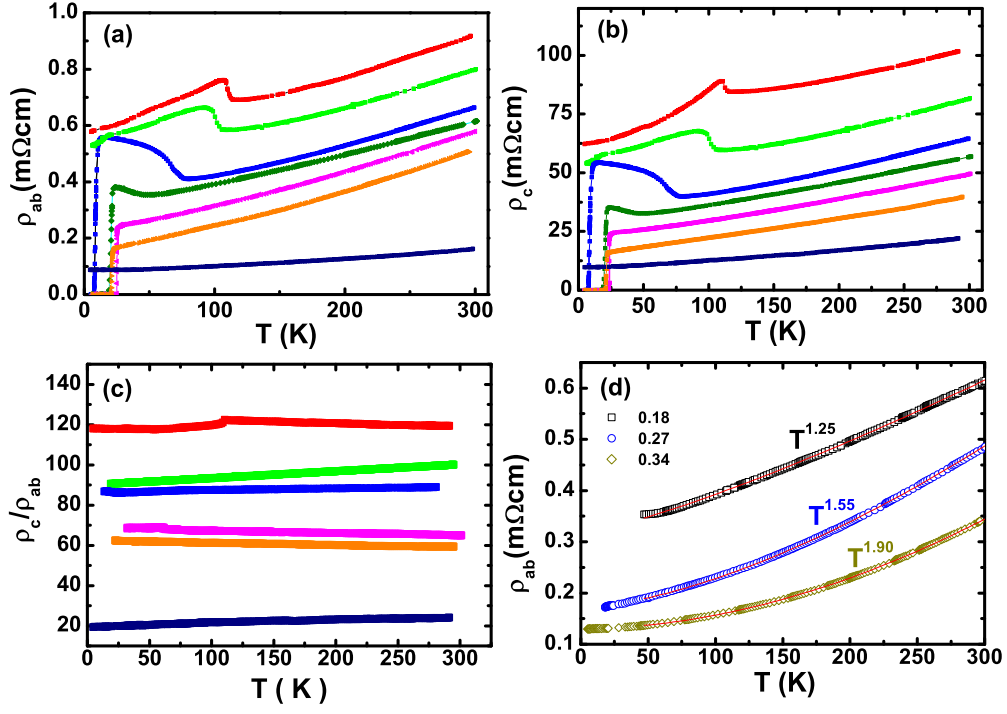}
\caption{Temperature dependent resistivity for the
$BaFe_{2-x}Co_{x}As_{2}$ single crystals. (a): In-plane resistivity
( 0.08-red , 0.12-green , 0.17-blue , 0.18-cyan , 0.20-magenta ,
0.25-yellow , 0.60-pink, respectively; (b): out-of-plane
resistivity; (c): Temperature-dependent anisotropy; (d): power law
fitting for x=0.18, 0.27 and 0.34. The red solid line is the fitting
line.}
\end{figure*}

\begin{figure*}[t]
\includegraphics[width=15cm]{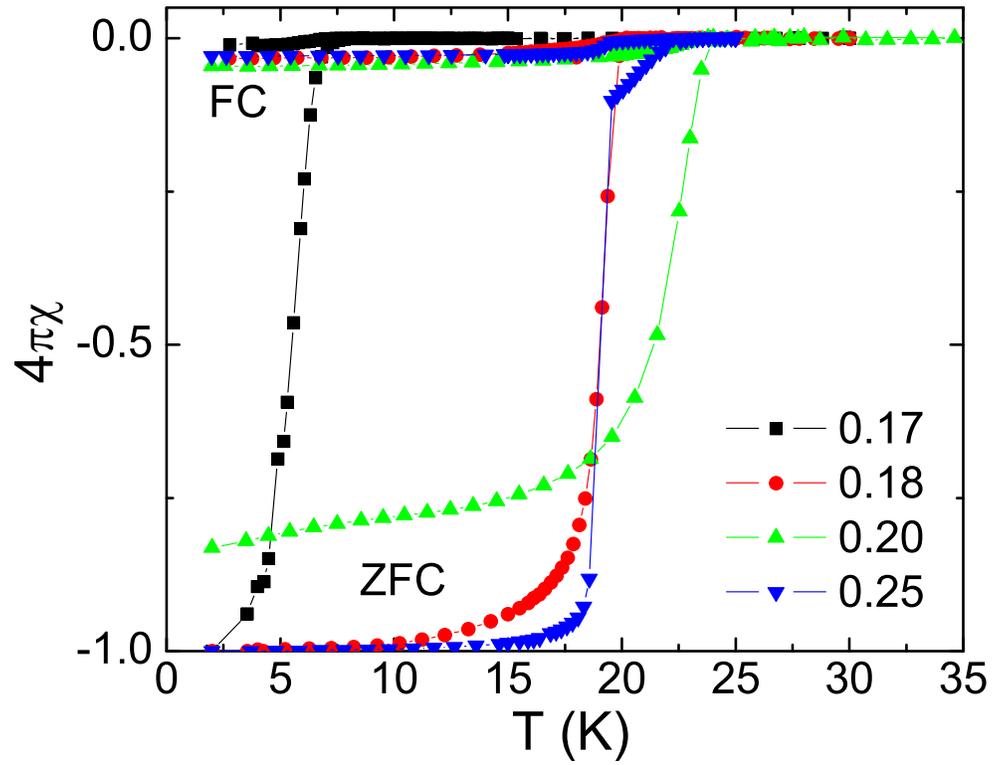}
\caption{Temperature dependent susceptibility for
BaFe$_{2-x}$Co$_{x}$As$_{2}$ (x=0.17, 0.18 ,0.20, 0.25) with
magnetic field 5 Oe.}\label{Fig:Fig5}
\end{figure*}

\begin{figure*}[t]
\includegraphics[width=12cm]{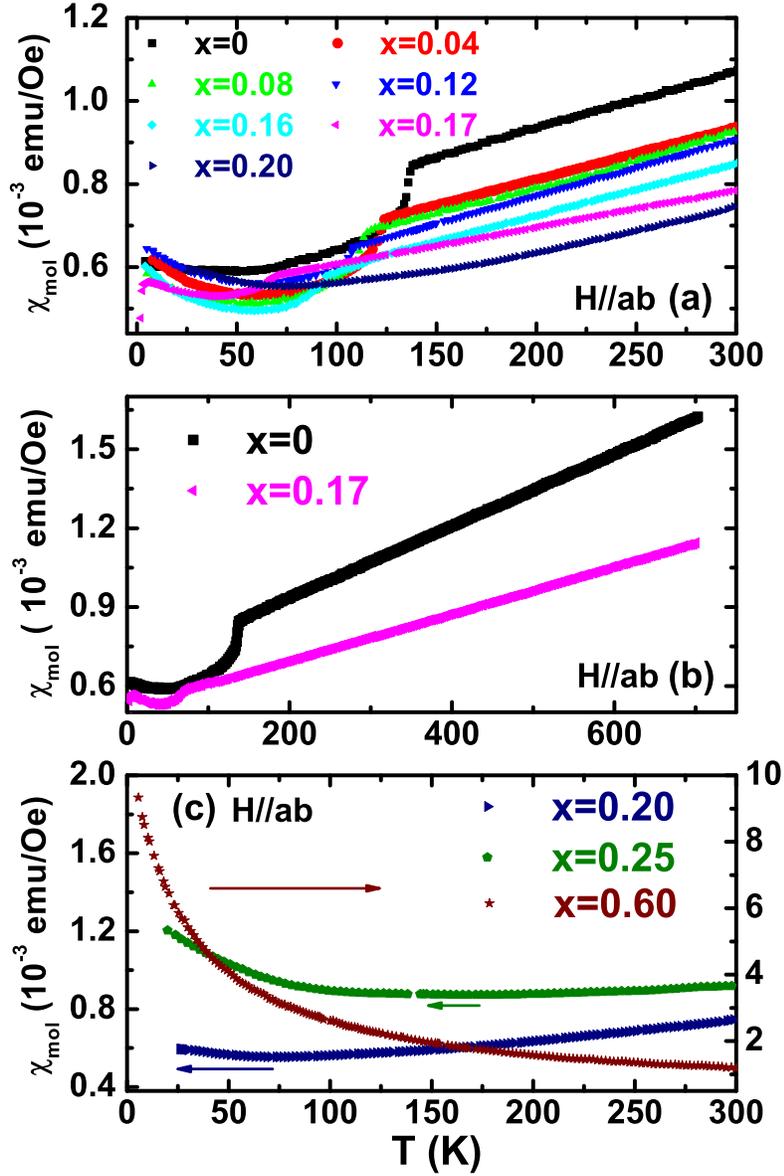}
\caption{Temperature dependent in-plane susceptibility for BaFe$_
{2-x}$Co$_{x}$As$_{2}$ single crystals under H =6.5T. (a): in the
temperature range from 2 K to 300 K for x$\leq$0.2; (b): in the
temperature range from 2 K to 700 K for x=0 and 0.17; (c): in the
temperature range from 2 K to 300 K for x=0.20,0.25 and
0.60.}\label{Fig:Fig4}
\end{figure*}

\begin{figure*}[t]
\includegraphics[width=15cm]{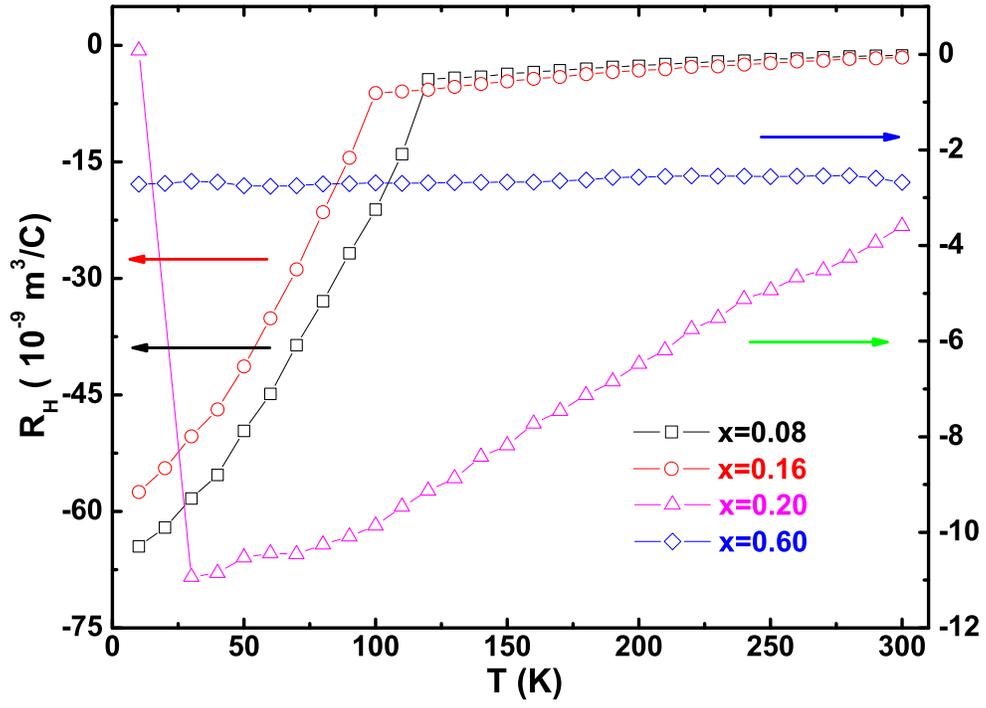}
\caption{Temperature dependent Hall coefficient for single crystals
BaFe$_ {2-x}$Co$_{x}$As$_{2}$ (x=0.08, 0.16, 0.20 and 0.60).
}\label{Fig:Fig7}
\end{figure*}

\begin{figure*}[t]
\includegraphics[width=15cm]{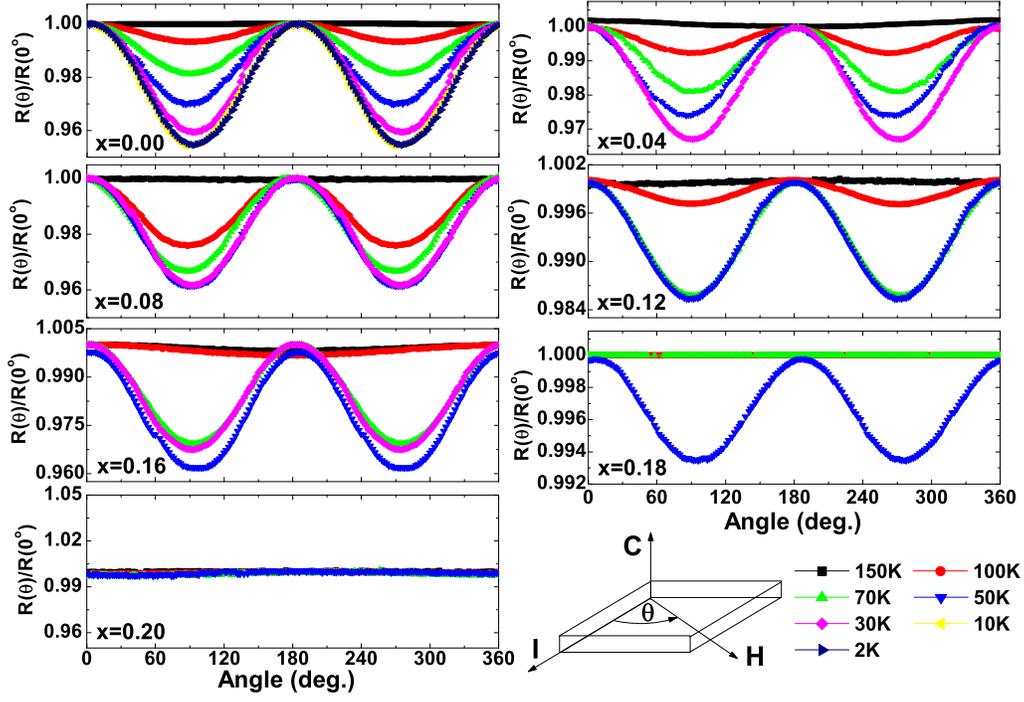}
\caption{Isothermal in-plane magnetoresistivity for $BaFe_
{2-x}Co_{x}As_{2}$ single crystals with rotating H within ab plane.
The measuring magnetic field is 14 T.}\label{Fig:Fig3}
\end{figure*}

\begin{figure*}[t]
\includegraphics[width=12cm]{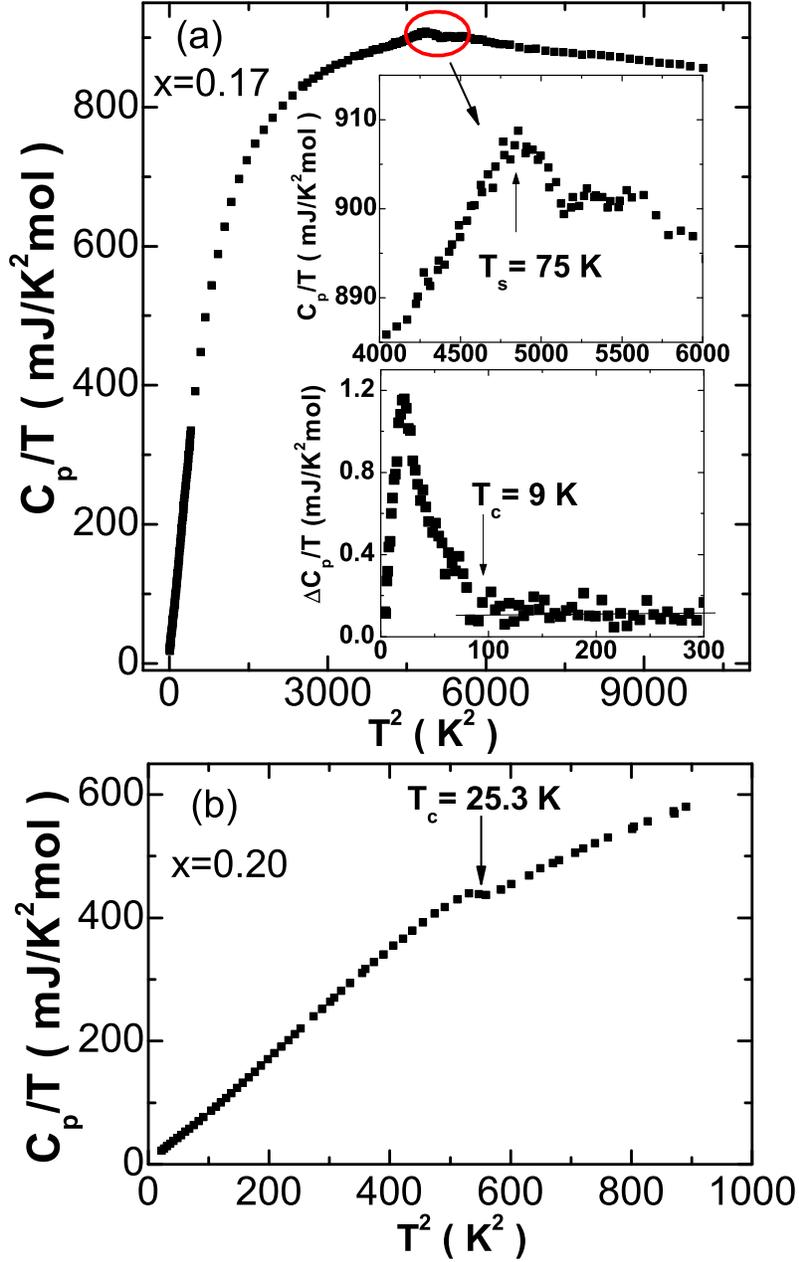}
\caption{Temperature dependent specific heat for $BaFe_
{2-x}Co_{x}As_{2}$ single crystals with x=0.17 and 0.20). (a):
$C_{p}/T$ vs $T^{2}$ for the crystal with x = 0.17. Up-inset: the
jump of specific heat around SDW transition; Down-inset: the jump of
specific heat around superconducting transition,
$\triangle$$C_{p}/T$ is obtained by substracting the heat capacity
of sample with low superconducting volume. (b): specific heat for
x=0.20. }\label{Fig:Fig8}
\end{figure*}

\begin{figure*}[t]
\includegraphics[width=15cm]{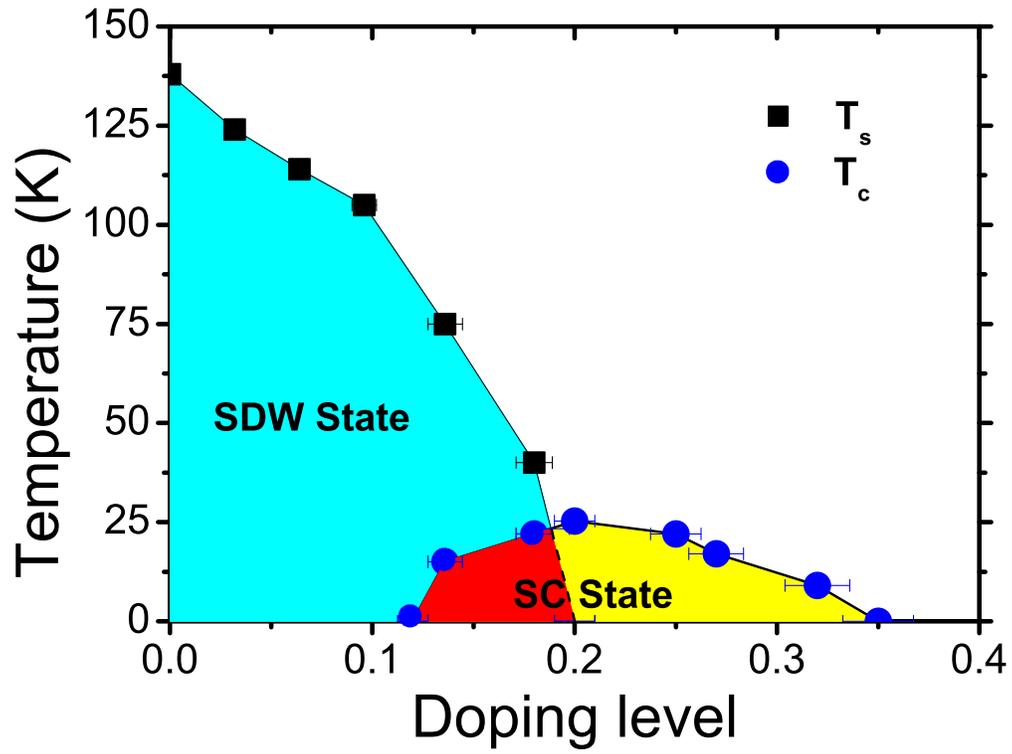}
\caption{Phase diagram of BaFe$_ {2-x}$Co$_{x}$As$_{2}$ within the
range $0{\leqslant}x{\leqslant }0.40$. Both Ts and Tc are
determined by resistivity.}\label{Fig:Fig6}
\end{figure*}

\end{document}